\documentstyle[11pt,paspconf,epsf]{article}
\begin{document}

\title{The X-ray outbursts of Be/X-ray transients}
\author{Ignacio Negueruela}
\affil{SAX Science Data Center, ASI, c/o Nuova Telespazio,
via Corcolle 19, I00131 Rome, Italy}
\author{Atsuo~T.~Okazaki}
\affil{Faculty of Engineering, Hokkai-Gakuen University,Toyohira-ku, Sapporo
062-8605, Japan }

\begin{abstract}
We present a new scenario for the behaviour of Be/X-ray binaries based on
long-term multiwavelength monitoring and the decretion disc model. 
The circumstellar discs of the primaries are truncated because of the
 tidal and resonant effect of the neutron
star. The geometry of the systems and the value
of viscosity determine the presence or absence of Type I X-ray outbursts.
The interaction of a strongly disturbed
disc with the neutron star originates Type II X-ray and optical outbursts. 
\end{abstract}


\section{Introduction}

Be/X-ray binaries are composed of a neutron star orbiting a Be star and 
accreting from its circumstellar disc. The high-energy radiation is believed 
to arise due to accretion of material 
associated with the Be star by the compact object (see Negueruela 1998; see
also Bildsten et al. 1997). Some Be/X-ray binaries are persistent X-ray 
sources (see Reig \& Roche 1999), 
displaying low-luminosity ($L_{{\rm x}} \sim 
10^{34}\:{\rm erg}\,{\rm s}^{-1}$) at
a relatively constant level (varying by up to a factor of $\sim 10$). On
the other hand, most known Be/X-ray binaries (though this is probably a 
selection effect) undergo periods in which the X-ray luminosity suddenly 
increases by a factor $\ga 10$ and are termed Be/X-ray transients. The 
distinction between the two groups is difficult to establish solely on 
terms of the temporal behaviour, since some sources such as A\,1118$-$61
and 4U\,1145$-$619 display relatively weak outbursts, though there is some
evidence that the X-ray spectral properties of the two groups could be
different.

Be/X-ray transients fall along a relatively narrow area in the 
$P_{{\rm orb}}$/$P_{{\rm spin}}$ diagramme (see Corbet 1986; Waters
\& van Kerkwijk 1989), indicating that some mechanism must be responsible
for the correlation. Those systems with fast-spinning neutron stars do not
show pulsed X-ray emission during quiescence (though non-pulsed radiation 
could be caused by accretion on to the magnetosphere) because of the
centrifugal inhibition of accretion (Stella et al. 1986). Systems with 
more slowly rotating pulsars show X-ray emission at a level $L_{{\rm x}} 
\la 10^{35}$ erg s$^{-1}$ when in quiescence. Transients show two different
kinds of outbursts:

\begin{itemize}
\item Moderate intensity X-ray outbursts ($L_{{\rm x}} 
\approx 10^{36} - 10^{37}$ erg s$^{-1}$) occurring in series separated by 
the orbital period (Type 
I or normal), generally (but not always) close to the time of 
periastron passage of the neutron star.  The duration of these 
outbursts seems to be related to the orbital period.
\item Giant (or Type II) X-ray outbursts ($L_{{\rm x}} 
\ga 10^{37}$ erg s$^{-1}$), lasting several weeks. The parameters
of these outbursts do not correlate clearly with
 orbital parameters, though in A\,0535+26 and 4U\,0115+63 they seem to
start always a few days after periastron passage. 
\end{itemize}

\section{Radial outflows vs. quasi-Keplerian discs}

Waters et al. (1989) tried to model the X-ray luminosities of Be/X-ray 
transients during
outbursts making  use of a simple wind accretion model, in which the 
neutron star accretes from a relatively fast radial outflow. The density 
in the disc of the Be primary was assumed to follow a power law,
as in Waters (1986) model for Be stars. In this scenario, the most relevant
parameter is the relative velocity between the outflow and
the neutron star, since the X-ray luminosity can then be expressed as
\begin{equation}
L_{{\rm x}} = 4\pi G^{3}M^{3}_{{\rm x}}R^{-1}_{{\rm x}}
v^{-4}_{{\rm rel}}F_{m} \propto \rho v^{-3}_{{\rm rel}}
\end{equation}
where  $M_{{\rm x}}$ and $R_{{\rm x}}$ are the mass and radius of the 
neutron star and $F_{m}=\rho v_{{\rm rel}}$ is the mass flow.
In order to explain the wide range of observed X-ray luminosities, large
changes in the value of the radial velocity have to be invoked. For 
example, Waters et al. (1989) deduced that the relative velocity was
$v_{{\rm rel}} \approx 300\:{\rm km}\,{\rm s}^{-1}$ during a Type I
outburst of V\,0332+53 in 1983, while it was $\ll 100\:{\rm km}\,{\rm s}^{-1}$
during a Type II outburst in 1973.

There is a large number of implicit assumptions in this formulation,
some of which are difficult to justify, but two obvious problems stand up. 
The first one is the low-luminosity X-ray emission displayed by many 
Be/X-ray transients when they are not in
outburst (for example, several detections of A\,0535+26 at
luminosities of $\approx 2\times10^{35}\:{\rm erg}\,{\rm s}^{-1}$ by
Motch et al. 1991). The model does not offer any explanation as to 
why there could be a change from quiescence to outburst, unless very 
large and sudden changes in the density and velocity of the flow are assumed,
while optical and infrared observations do not show any sign of the 
large variations that would be associated with a change of several orders of
magnitude in the density of material. 

The major objection to the model is simply the fact that there 
is no observational evidence whatsoever supporting the 
existence of such fast outflows. All observations of Be discs imply 
bulk outflow velocities smaller than a few ${\rm km}\,{\rm s}^{-1}$.
The evidence for rotationally dominated quasi-Keplerian discs around Be 
stars is overwhelming (see Hanuschik et al. 1995; Hummel \& Hanuschik 1997), 
specially due to the success of the one-armed global
oscillation model to explain V/R variability in the emission lines of
Be stars (see Okazaki 1991, 1997; Papaloizou et al. 1992; Hummel \& Hanuschik 
1997).

The discovery by Reig et al. (1997) of a correlation between the
maximum equivalent width reached by the H$\alpha$ emission line in 
Be/X-ray binaries and their orbital period strongly suggested that 
the neutron star had some kind of effect on the disc of the Be primary. 
Other observational facts, such as circumstantial evidence for the
discs in Be/X-ray binaries being optically thicker than those of
isolated Be stars, added further support to this idea.

This, together with the increasing evidence for quasi-Keplerian discs 
around Be stars, has prompted us to investigate whether the properties
of Be/X-ray binaries can be better explained if we assume that the 
disc surrounding the Be primary is a viscous decretion disc. Due to space
limitations, we will not try to argue the case here, but refer the
interested reader to future publications (Negueruela \& Okazaki 2000;
Negueruela \& Okazaki, in preparation) and concentrate on the derived
model. We just note that the choice of the 
viscous decretion disc model has been prompted by its success at 
explaining observational characteristics of Be stars (Lee et al. 1991;
Porter 1999; Okazaki 2001), but
it is not a necessary condition for the majority of the conclusions 
below. As a matter of fact, the tidal and resonant interaction of the
neutron star can effectively truncate the Be star disc for any model 
in which the outflow velocity in the disc is subsonic.

\section{Disc truncation}

A viscous decretion disc is held by the outwards diffusion of angular
momentum due to viscous interaction. This viscous torque communicates
angular momentum to the outflowing material, allowing it to follow
quasi-Keplerian orbits. A neutron star orbiting the Be star exerts a
negative torque on the material which takes away angular momentum. The 
competing effects of viscosity and resonant torques determine the radial
distance at which the disc will be truncated. The effect of the neutron
star is most strongly felt at the $n:1$ commensurabilities between
the orbits of the neutron star and material at that distance, and 
truncation is likely to take place there. For every resonance $n:1$,
there is a critical value of the viscosity parameter, $\alpha_{\rm crit}$,
such that  if $\alpha < \alpha_{\rm crit}$, the disc is truncated at
that resonance. The value of  $\alpha_{\rm crit}$ depends on the orbital
parameters of the system under consideration.

\begin{figure}[t]
\plotfiddle{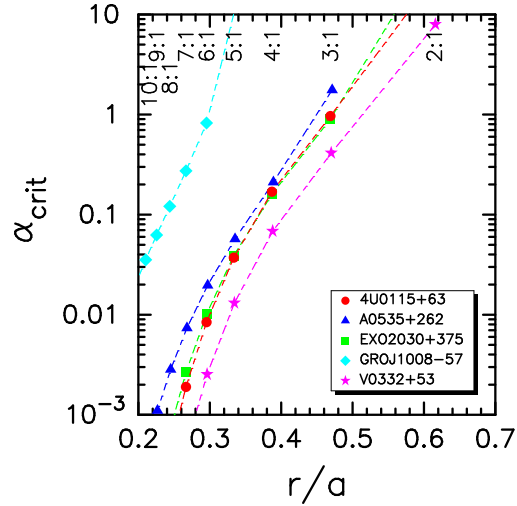}{6cm}{0}{40}{40}{-120}{-24}
\caption{Values of $\alpha_{\mathrm crit}$ for several Be/X-ray binaries.
The lower axis shows the distance from the central star (normalised
to the mean orbital separation $a$) at which truncation
will occur for a given viscosity. The position of the $n:1$ 
commensurabilities is indicated under the upper axis.}
\label{fig:systems}
\end{figure}

Fig.~\ref{fig:systems} shows the values of $\alpha_{\rm crit}$ that will
cause truncation at a given resonance radius for a number
of Be/X-ray binaries. The systems have been modelled using as masses for 
the Be primaries those corresponding to their spectral type, when accurately
determined. GRO J1008$-$57 and EXO\,2030+375, were assumed
to have spectral type B0V and mass $M_{*}=18\,M_{\sun}$. The orbital
parameters are those determined from the analysis of Doppler-shift in the 
pulse arrival time during X-ray observations. Assuming reasonable
values for the viscosity ($\alpha \sim 0.1$), we find that all systems, except
GRO J1008$-$57, will be truncated at either the $4:1$ or $3:1$ resonance
radius. In all
five systems, the likely truncation radii are very close
to the size of the effective Roche radius at periastron. 

\begin{figure}[t]
\plotfiddle{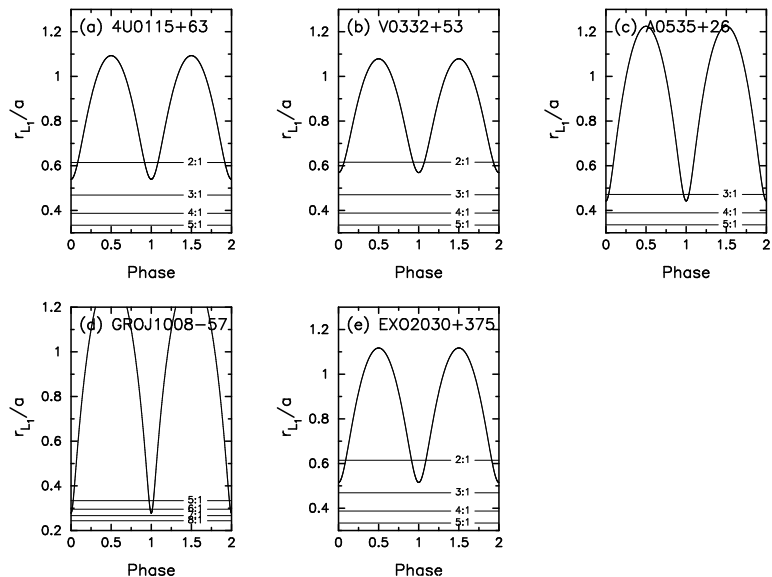}{8.5cm}{0}{100}{100}{-288}{-180}
\caption{Some orbital parameters for the Be/X-ray binaries considered
in the text.
The distance of the first Lagrangian point $L_{1}$ (normalised
to the mean orbital separation $a$) is plotted against
orbital phase. The position of the $n:1$ 
commensurabilities is indicated by the horizontal lines.}
\label{fig:vcas}
\end{figure}

For the systems with close orbits (4U\,0115+63, V\,0332+53
and EXO\,2030+375), the
first Lagrangian point $L_{1}$ is always further away than the $3:1$ 
resonance (see Figure~\ref{fig:vcas}). In A\,0535+26, on the other hand,
$r(L_{1}) < r(3:1)$ close to periastron, while $r(L_{1}) > r(4:1)$
always. Similarly, for GRO J1008$-$57 $r(L_{1}) < r(6:1)$ close to 
periastron, while $r(L_{1}) > r(7:1)$ always.

The scenario suggested by these results is as follows. For close systems,
the truncation of the disc prevents the accretion of
significant amounts of material by the neutron star. Even though the 
truncation is not expected to be 100\% effective, centrifugal inhibition
of accretion for very low accretion rates (Stella et al. 1986) leads
to the absence of quiescence X-ray luminosity. However, these systems
can still present series of Type I outbursts if the discs surrounding
the central Be stars are very disturbed. If some sort of perturbation,
such as a global density wave (Okazaki 1997) or radiation-induced warping 
(Porter 1998), produces an eccentric disc, then material can be accreted 
through $L_{1}$ close to periastron. This situation will lead to short and
irregular series of X-ray outbursts, very likely showing decreasing
intensity as in the 1996 series of outbursts of 4U\,0115+63 
(Negueruela et al. 1998).

Systems with wider orbits, such as A\,0535+26 and GRO J1008$-$57 
can show series of Type I
outbursts if the viscosity is high enough to allow the disc to extend
beyond $r(L_{1})$. Small changes in the viscosity can result in the system
switching on or off for relatively long periods. The outbursts in one
series will be of approximately the same strength, unless some other
effects are involved (it is not clear, for example, how a mediating 
accretion disc could affect the mass transfer).

\begin{table}[t]
\caption{Orbital and stellar parameters used in the modelling of 
the Be/X-ray binaries considered. For references, see Negueruela (1998).
The orbital parameters for GRO J1008$-$57 are based on the best fit
to the BATSE data (M. Scott, priv. comm.). Data marked with a `*' are
assumed values.} 
\label{tbl-1}
\begin{center}
\begin{tabular}{lcccccc}
Name & $P_{{\rm s}}$(s) & $P_{{\rm orb}}$(d) & $e$ & Optical & Spectral & Mass\\
& & & & Counterpart & Type & ($M_{\sun}$) \\
\tableline
4U\,0115+63 &3.6 & 24.3 & 0.34 &V635 Cas & B0.2V & 18\\
V\,0332+53 & 4.4 & 34.2 & 0.31 & BQ Cam & O8.5V & 20 \\
A\,0535+26 & 103.5 & 110.3 & 0.47 & V725 Tau & O9.7III & 23 \\
GRO J1008$-$57 & 93.5 & 247.5 & 0.66 & star & B0V* & 18*\\
EXO\,2030+375 & 41.7 & 46.0 & 0.36 & star & B0V* & 18*\\
\end{tabular}
\end{center}
\end{table}

The scenario seems to reproduce well the observed properties of most
of the systems. Both 4U\,0115+63 and V\,0332+53 show extended periods
of quiescence during which no X-ray emission at all is detected. 
V\,0332+53 seems to have kept a very small disc for the last $\sim 7$
years (Negueruela et al. 1999). The outer rim of the H$\alpha$ emitting
region, as determined from the peak separation, is at a distance similar
to our calculation for the $4:1$ resonance.  4U\,0115+63 has only once 
displayed a series of Type I outbursts in $\sim 30$ years of observations
and this was when the disc of the Be star was very disturbed (Negueruela
et al. 1998). V\,0332+53 has only displayed a series of three Type I
outbursts since its discovery. Another source with comparable behaviour is
2S\,1417$-$62, which has similar orbital parameters.

On the other hand, A\,0535+26 shows longer series of Type I outbursts as 
well as long periods of quiescence (during which it displays low-luminosity
X-ray emission). The change between these two states can be explained as
a consequence of small variations in the physical conditions in the disc
which result in the outer edge moving between the $3:1$ and $4:1$ 
resonances. 

It must be pointed out that the behaviour of EXO\,2030+375 is very different
from that predicted, since it displays very long series of Type I outbursts,
similar to those predicted for systems with larger eccentricities. In
principle, it could be argued that this is an indication of very high
viscosity ($\alpha \sim 1$), but we note that there is no observational
knowledge about the stellar parameters of the central star, which is
too heavily obscured to allow even an approximate spectral classification.
The actual values for the mass and radius of the Be star could be very
different from those used here.

\section{Disc dynamics and Type II outbursts}

An interesting consequence of the model presented above is the fact that
the discs surrounding the Be primaries cannot reach a steady state. Material,
having lost the angular momentum needed to outflow, will fall back and
form a dense torus close to the truncation resonance. This situation
will lead to the dynamical instability of the disc, which will produce
large-scale perturbations, such as warping or global density waves. Such
perturbations will either lead to the dispersion of the disc or settle
down after some time. Therefore we expect the discs to undergo cycles of 
reformation and dissipation and/or major perturbations and resettling.

\begin{figure}[t]
\plotfiddle{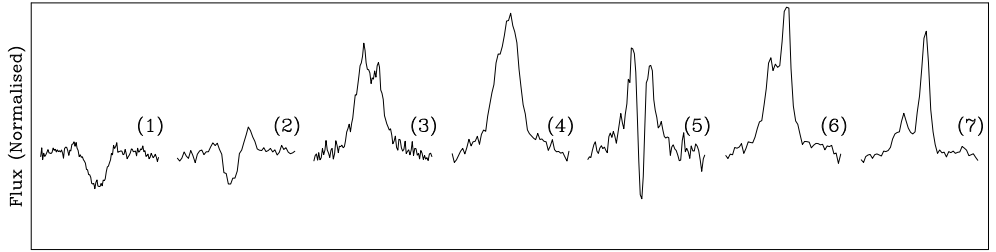}{4cm}{0}{95}{95}{-288}{-216}
\caption{A hypothetical cycle for the H$\alpha$ line in V635 Cas, the 
optical counterpart to 4U\,0115+63, created by superposing the observations
from 1997\,--\,1998 (profiles 1\,--\,4; present cycle) and those
from 1995\,--\,1996 (profiles 5\,--\,7; end of previous cycle).
The cycle starts with the absence of the circumstellar disc (profile 1).
When the disc appears (2), it quickly grows into a typical double-peaked 
profile (3). Gradually,
the peaks converge until a single-peak profile is seen (4), indicating the
warping of the disc. This is followed by a number of fast transitions
between single-peaked and shell profiles (5), as the warped disc precesses.
Finally, the disc is very perturbed and the asymmetry grows as the strength
of the emission line decreases (6 \& 7), leading to a new disc-less phase.}
\label{fig:cycle}
\end{figure}

 In the case of V635 Cas, the optical counterpart to 4U\,0115+63, 
spectroscopic monitoring has shown that the Be star undergoes cycles
of disc reformation, warping and dissipation with time-scales of 3\,--\,5
years (see Figure~\ref{fig:cycle}). These cycles are reflected in the 
X-ray behaviour of the source, which has shown quasi-periodicities of
$\sim 3$ years, associated with the main cycle, and $\sim 6\,-\,8$ months,
when the perturbed disc is precessing. BQ Cam, the optical counterpart
to V\,0332+53, on the other hand, suffered some major perturbation 
around the time of its last Type II outburst in 1989, but then settled
down to a semi-steady state (Negueruela et al. 1999). Unfortunately, 
monitoring of the optical counterparts to other sources has been very
sparse. 

The large perturbation in the disc of V725 Tau, the optical
counterpart to A\,0535+26, which occurred at the time of its last
Type II outbursts (Negueruela et al. 1998) has led to the gradual
dissipation of the circumstellar envelope (see Haigh et al. in this
proceedings). Such behaviour is reminiscent of that observed in
V635 Cas and is probably indicating some similarly quasi-cyclical
activity on a longer time-scale.

The association of large-scale perturbations and Type II outbursts in
Be/X-ray binaries had already been noted by Negueruela et al. (1998).
The observations of V635 Cas showing very fast changes in the shape
and width of the emission lines indicate that the circumstellar disc
of this source can become warped and precess. Observations from 1989 (covering
the last three cycles of disc loss and reformation) indicate that
X-ray activity occurs only after the disc is warped, showing that
it is the perturbation in the disc which leads to the outburst, and not
otherwise. This is perfectly consistent with the model outlined above. 
Estimates of the accretion rates during Type II outbursts show that 
a significant fraction of the disc material has to be accreted by
the neutron star. Since the disc is truncated, accretion of large amounts
of matter will only be possible if the disc has become sufficiently 
asymmetric and dense to overflow the truncation radius.

\section{The global view}

The main advantage of the model outlined in this paper over previous ones
is that it provides a global picture, though still a very sketchy one, in
which the whole phenomenology of Be/X-ray binaries is seen as deriving from
a small set of simple physical facts. The truncation of the discs 
surrounding the Be stars by the neutron star companions provides an 
explanation for the long periods of quiescence, while the dependence of
the truncation radius on the physical properties of the disc gives a 
natural way of understanding the onset of the series of Type I outbursts
and their eventual disappearance. Our modelling of different systems 
shows that in all cases the parameters involved are such that values
of the viscosity in the range expected
from theoretical considerations and modelling  ($0.01 < \alpha < 1$)
result in truncation at distances comparable to the size of the effective
Roche lobe of the Be star. Moreover, in all cases, small changes of
the viscosity result in variations in the distance at which truncation
takes place.

As a consequence of disc truncation, the circumstellar discs cannot be
steady, which will lead to the development of the perturbations that have
been observed. When the perturbations give rise to large asymmetries in
the density distribution, the truncation mechanism will be much less 
effective and large amounts of material will be able to make their way
to the neutron star, producing the giant Type II outbursts.

\acknowledgments
IN is supported by an ESA external fellowship.


\begin{references}

\reference Bildsten, L., Chakrabarty, D., Chiu, J., et al. 1997, ApJS,
113, 367

\reference Corbet, R.H.D. 1986, MNRAS 220, 1047

\reference Hanuschik, R.W., Hummel, W., Dietle, O., \& Sutorius, E., 
1995, A\&A, 300, 163 

\reference Hummel, W., Hanuschik, R.W. 1997, A\&A, 320, 852

\reference Lee, U., Saio, H, Osaki, Y. 1991, MNRAS, 250, 432

\reference Motch, C., Stella, L., Janot-Pacheco, E., \& Mouchet, M.
1991, ApJ, 369, 490

\reference Negueruela,  I. 1998, A\&A, 338, 505

\reference Negueruela,  I., \& Okazaki, A.T. 2000, A\&A, accepted

\reference Negueruela, I., Reig, P., Coe, M.J., \& Fabregat, J. 
1998, A\&A, 336, 251

\reference Negueruela, I., Roche, P., Fabregat, J.,\& Coe, M.J. 
1999, MNRAS, 307, 695

\reference Okazaki, A.T. 1991, PASJ, 43, 75

\reference Okazaki, A.T. 1997, A\&A, 318, 548

\reference Okazaki, A.T. 2001, PASJ, in press 

\reference Papaloizou, J.C., Savonije, G.J., \& Henrichs, H.F. 1992,
A\&A, 265, L45

\reference Porter, J.M. 1998, A\&A, 336, 966

\reference Porter, J.M. 1999, A\&A, 348, 512

\reference Reig, P., \& Roche, P. 1999, MNRAS, 306, 100

\reference Reig, P., Fabregat, J., \& Coe M.J.,
1997, A\&A, 322, 193

\reference Stella, L., White, N.E., \& Rosner, R. 1986, ApJ, 308, 669

\reference Waters, L.B.F.M. 1986, A\&A, 162, 121

\reference Waters, L.B.F.M., \& van Kerkwijk, M.H. 1989, A\&A, 223, 196

\reference Waters, L.B.F.M., de Martino, D., Habets, G.M.H.J., 
\& Taylor, A.R. 1989, A\&A, 223, 207

\end{references}
\end{document}